# Real-Time Data Mining of Massive Data Streams from Synoptic Sky Surveys

S. G. Djorgovski, M. J. Graham, C. Donalek, A. A. Mahabal, A. J. Drake

California Institute of Technology, Pasadena, CA 91125, USA
Emails: [djorgovski,mjg,donalek,aam,ajd]@cd3.caltech.edu

M. Turmon, T. Fuchs[*]

Jet Propulsion Laboratory, California Institute of Technology, Pasadena, CA 91109, USA
Emails: [turmon, thomas.fuchs]@jpl.nasa.gov
* Present address: Memorial Sloan Kettering Cancer Center, New York, NY 10065

HIGHLIGHTS
___________________________________________________________________________

- Advances in the automated classification of transient events in synoptic sky surveys
- Innovative methods for the analysis of irregularly sampled, heterogeneous time series
- Novel approach to the machine-assisted discovery using a symbolic regression
- Approaches to an automated decision making based on the automated classification

___________________________________________________________________________

ABSTRACT
___________________________________________________________________________

The nature of scientific and technological data collection is evolving rapidly: data volumes and rates grow exponentially, with increasing complexity and information content, and there has been a transition from static data sets to data streams that must be analyzed in real time. Interesting or anomalous phenomena must be quickly characterized and followed up with additional measurements via optimal deployment of limited assets. Modern astronomy presents a variety of such phenomena in the form of transient events in digital synoptic sky surveys, including cosmic explosions (supernovae, gamma ray bursts), relativistic phenomena (black hole formation, jets), potentially hazardous asteroids, etc. We have been developing a set of machine learning tools to detect, classify and plan a response to transient events for astronomy applications, using the Catalina Real-time Transient Survey (CRTS) as a scientific and methodological testbed. The ability to respond rapidly to the potentially most interesting events is a key bottleneck that limits the scientific returns from the current and anticipated synoptic sky surveys. Similar challenge arise in other contexts, from environmental monitoring using sensor networks to autonomous spacecraft systems. Given the exponential growth of data rates, and the time-critical response, we need a fully automated and robust approach. We describe the results obtained to date, and the possible future developments.
___________________________________________________________________________

## I. INTRODUCTION

The scientific measurement and discovery process traditionally follows the pattern of theory followed by experiment, analysis of results, and then follow-up experiments, often on time scales from days to decades after the original measurements, feeding back to a new theoretical understanding. But that clearly would not work in the case of phenomena where a rapid change occurs on time scales shorter than what it takes to set up the new round of measurements. Thus there is a need for autonomous, real-time scientific measurement systems, consisting of discovery instruments or sensors, a real-time computational analysis and decision engine, and optimized follow-up instruments that can be deployed selectively in (or in near) real-time, where measurements feed back into the analysis immediately. The need for a rapidly analysis, coupled with massive and persistent data streams, implies *a need for an automated classification and decision making*.

This entails some special challenges beyond traditional automated classification approaches, which are usually done in some feature vector space, with an abundance of self-contained data derived from homogeneous measurements. The input information here is generally sparse and heterogeneous: there are only a few initial measurements, their types differ from case to case, and the values have differing variances; the contextual information is often essential, and yet difficult to capture and incorporate; many sources of noise, instrumental glitches, etc., can masquerade as transient events; as new data arrive, the classification must be iterated dynamically. There is also the requirement of a high completeness (don't miss any interesting events) and low contamination (not too many false alarms), and the need to complete the classification process and make an optimal decision about expending valuable follow-up resources (e.g., obtain additional measurements using a more powerful instrument, diverting it from other tasks) in real time. These challenges require novel approaches.

Astronomy in particular is facing these challenges in the context of the rapidly growing field of time domain astronomy, based on the new generation of digital synoptic sky surveys that cover large areas of the sky repeatedly, looking for sources that change position (e.g., potentially hazardous asteroids) or change in brightness (a vast variety of variable stars, cosmic explosions, accreting black holes, etc.). Time domain touches upon all subfields of astronomy, from the Solar system to cosmology, and from stellar evolution to the measurements of dark energy and extreme relativistic phenomena. Many important phenomena can be studied only in the time domain (e.g., Supernovae or other types of cosmic explosions), and there is a real possibility of discovering some new, previously unknown types of objects or phenomena.

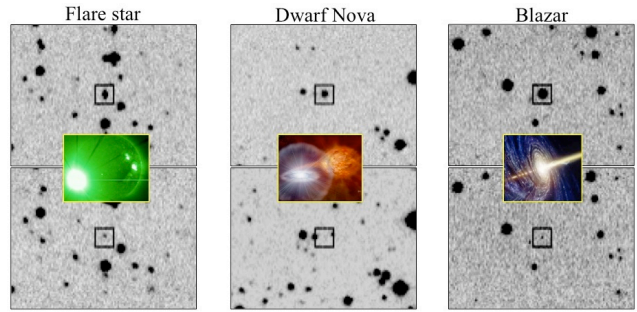

**Figure 1.** An illustration of the classification challenge, using examples of transient events from the Catalina Real-time Transient Survey (CRTS) [20,21,22,23,24]. Images in the top row show objects which appear much brighter that night, relative to the baseline images obtained earlier (bottom row). On this basis alone, the three transients are physically indistinguishable, yet the subsequent follow-up shows them to be three vastly different types of phenomena: a flare star (left), a cataclysmic variable powered by an accretion to a compact stellar remnant (middle), and a blazar, flaring due to instabilities in a relativistic jet (right). Accurate transient event classification is the key to their follow-up and physical understanding.

However, while the surveys discover transient or variable sources, the scientific returns are in their physical interpretation and follow-up observations (Fig. 1). This entails physical classification of objects on the basis of the available data, and an intelligent allocation of limited follow-up resources (e.g., time on other telescopes or space observatories), since generally only a small fraction of all detected events can be followed, and some of them are much more interesting than others. Large data rates and the need for a consistent response imply the need for the automation of these processes, and the problem is rapidly becoming much worse. Today, we deal with data streams of the order of ~ 0.1 TB/night and some tens of transients per night; the upcoming Large Synoptic Survey Telescope (LSST) [1] is expected to generate ~ 20 TB/night, and millions of transient event alerts. The planned Square Kilometer Array (SKA) [2] radio telescope will move us into the Exascale regime. Thus, a methodology for an automated classification and follow-up prioritization of transient events and variable sources is *critical for the maximum scientific returns* from these planned facilities, in addition to enabling the time domain science now.

In general, most of the major astronomical data sets today are connected and accessible through the Virtual Observatory (VO) framework, which is effectively the global data grid of astronomy [49,50,51]. However, VO so far does not incorporate many services for knowledge extraction from the massive data sets or data streams, and this is especially important in the context of the time domain astronomy.

The challenges stem from several reasons: first, the data are sparse, especially right after the initial detection; archival and contextual information is essential (e.g., the spatial context of the source, the multi-wavelength context, and the temporal context – has the source been detected before, and if so, what was its variability behavior, etc.). Both the subsequent measurements (if any) and the archival information are likely to be highly heterogeneous and/or incomplete. The probabilistic classification of the events evolves as new data arrive, and is used to generate priorities and automated follow-up decisions and requests, which are then feed back into the system. However, the availability of the follow-up resources also changes in time, which affects their value, and is limited in allocation; etc.

To respond to these challenges, we have been developing and testing a variety of automated classification approaches for time domain astronomy. We divided the problem into two parts: event classification, and follow-up recommendations given the available assets. Our preliminary results have been described, e.g., in [3,4,5,6,7,8,9,10,11,12,37]. Here we give some updates to these papers and some of our current work. For additional reviews and references, see, e.g., [13,14,15, 16,17,18,19].

As a testbed development data stream, we use transient events and variable sources discovered by the Catalina Real-Time Transient Survey (CRTS) [20,21,22,23,24]. CRTS provides a great variety of physical object types, and a realistic heterogeneity and sparsity of data. We found that a number of published methods, developed on "de luxe" data sets, to say nothing about the simulated data, simply fail or significantly underperform when applied to the more realistic data (in terms of the cadences, S/N, seasonal modulation, etc.), typified by the CRTS data stream. In general, we find that every method has some dependence on the quantity and quality of the input data (e.g., the number of measurements in a light curve, the sampling strategy, etc.), and all of our tests incorporate assessment of the robustness and applicability of a given method in different data regimes.

Whereas our focus is on an astronomical context, similar situations arise in may other fields, where anomalies or events of interest must be identified in some massive data stream, characterized, and responded to in as close to the real time as possible (e.g., environmental monitoring, security, etc.).

## II. BAYESIAN NETWORKS

Bayesian techniques may be the most promising approach for the classification with sparse, incomplete, or missing data, since, generally speaking, one can use the information from the available priors, regardless of what data are not available. In particular, we experimented with a Bayesian Network (BN) [25] based classifier, as it offers a natural way of incorporating a variety of the measurements of different types, and more can be added as they become available. However, the network complexity increases super-exponentially as more variables are included, and there is a premium of selecting a small number of the most powerful classification discriminating features (see below).

Our initial implementation used follow-up measurements of photometric colors obtained at the Palomar 60-inch telescope. For example, in the relative classification of Cataclysmic Variables (CVs) vs. Supernovae, we obtain a completeness of ~ 80% and a contamination of ~ 19%. Whereas that is a respectable performance for the current state of the art in this field, we were able to improve on it considerably, by including other, contextual information.

We found BN to be an excellent way of incorporating quantitative spatial contextual information, e.g., the proximity of a given transient event to the nearest star or the nearest galaxy detected in the Sloan Digital Sky Survey (SDSS) [26]. For example, a transient (nearly) coincident with a galaxy will most likely be a Supernova (SN), whereas a transient coincident to with a star-like object in an archival survey such as the SDSS would more likely be some type of a variable star or an Active Galactic Nucleus (AGN). Both of these are limited by the depth and the angular resolution of the comparison archival survey, but for our tests, transients from CRTS and comparisons with SDSS are well matched for this purpose.

In the case of proximity to the nearest star, a simple angular distance is sufficient. In the case of galaxies, an ambiguity arises: is a closer, but very faint galaxy more likely to be the possible SN host, or a considerably brighter galaxy that is a little further away? Thus, a different metric is needed, and we use angular separation in the units of characteristic radii for the light distribution in galaxies. After some experimentation, we decide on the so-called Petrosian radius, which is one of the parameters provided by the SDSS archive.

Temporal contextual information is also important. Another distinguishing characteristic of SNe is that they can explode only once, so a presence of previously detected spikes in the light curve of a given transient diminishes the likelihood of it being a SN.

Thus, we construct a BN with 3 input variables, the proximity to the nearest star, to the nearest galaxy (suitably normalized), and a light curve peak statistic developed by us. The results are illustrated in Fig. 2. The results using just these 3 contextual variables (the nearest star distance, the nearest galaxy distance, and the peak statistics) are very encouraging. For the transients correctly classified as SNe, the completeness is in the range ~ 80% – 92% with contamination in the range ~ 18% – 29%. For the transients correctly classified as *not* being SNe, the completeness is in the range ~ 79% – 83% with contamination in the range ~ 8% – 14%. Given the modest number of input variables, we find these results to be very encouraging.

The purpose of this experiment was mainly to evaluate the utility of BNs for the inclusion of contextual variables, such as the distance to the nearest star or a galaxy. We believe that these results can be improved substantially by introducing other priors, e.g., colors, or light curve based parameters; these experiments are still in progress. In general, the performance of BNs can be improved at the expense of an increased computational complexity. As we already noted, the number of Directed Acyclic Graphs (DAGs), of which BNs are a particular example, grows hyper-exponentially as additional nodes (variables) are added; see, e.g., [43,44]. This further emphasizes the problem of feature selection, which we address below.

III. STATISTICAL DESCRIPTORS OF VARIABILITY AND THE OPTIMAL FEATURE SELECTION

Data heterogeneity is perhaps the key problem for the automated classification of astronomical light curves, or, for that matter, any other irregularly sampled time series. Since the numbers of the data points and their temporal separations vary, the light curves themselves cannot be used directly in any method that assumes data in the form of uniform feature vectors. In order to circumvent this problem, we evaluate a number of statistical descriptors of light curves that can be evaluated regardless of the number of data points or the cadence, e.g., the variance of the observed magnitudes, the skew, kurtosis, etc. About 60 such parameters have been defined in the literature, to which we added a dozen of our own devising. Their definitions can be found at the *Caltech Time Series Characterization Service* [27]. These statistical descriptors can then be used to form feature vectors that can be fed into automated classifiers.

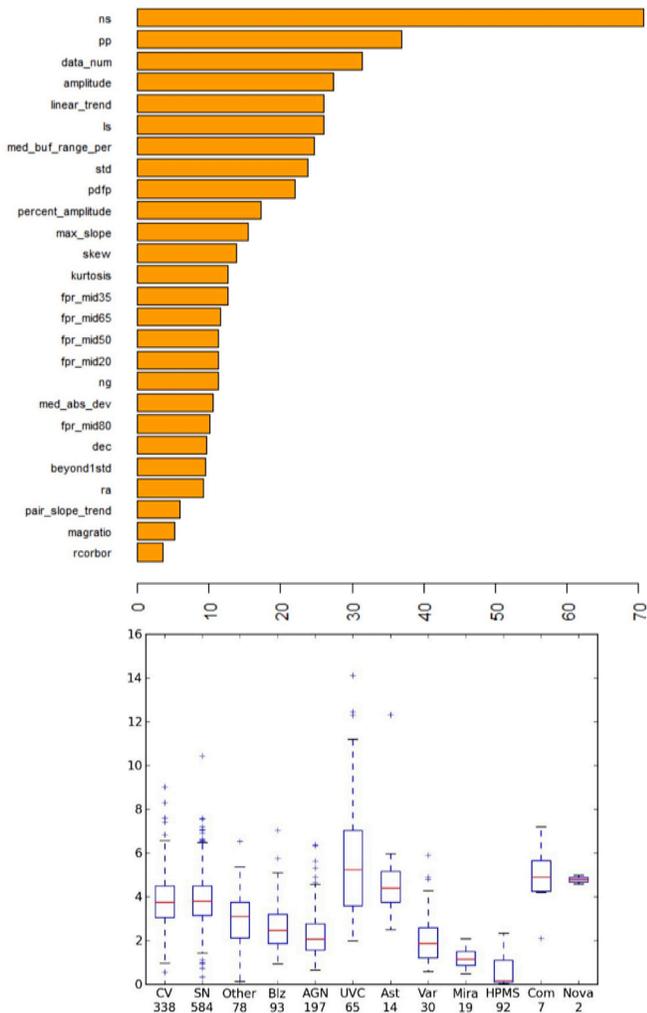

**Figure 3.** Top: A relative ranking of light curve features in terms of the classification discriminating power. For the definitions of these parameters, see [27]. Bottom: Standard box plot representation of the distributions of the top ranked parameter (*nsigma*) for different physical types of transients and variables.

We believe that this substitution, turning an irregular set of measurements for a set of time series (or indeed any other ordered, 1-dimensional set of measurements) into a homogeneous set of statistical descriptors that form complete feature vectors that can be fed into the standard machine classifiers that operate on feature vectors, can be very useful in many other domains or situations where the data are irregularly sampled, incomplete (missing measurements are simply ignored), or otherwise inhomogeneous. In principle, a well designed set of statistical descriptors would contain all of the information that is present in the data.

Obviously, not all would be equally useful, and different ones may be more useful in different circumstances. We have conducted a detailed study of their utility for different aspects of the classification problem, for different classifiers, and in different data regimes (e.g., S/N, number of data points, etc.) for the CRTS data set. Ideally, one seeks combinations of features that optimally separate different classes of transients or variables. Here we summarize some of the key results; more details are given in [28].

Given a set of feature vectors, a broad variety of automated classification tools can be applied, both supervised and unsupervised. Supervised methods include artificial neural networks (ANN), and in particular the multi-layer perceptron (MLP), support vector machines (SVM), decision trees (DT) and their generalization random forests (RF), etc. Unsupervised methods include Kohonen self-organizing maps (SOM), *k* Means (KM), *k* nearest neighbors (kNN), etc. Given a particular classifier, and a particular classification problem, e.g., separating two different types of periodic variables, or supernovae vs. non-explosive transients, we can evaluate the relative importance of different features using several methods.

One way to reduce the dimensionality of the input space is applying a forward feature selection strategy that consists in selecting a subset of features from the training set that best predict the test data by sequentially selecting features until there is no improvement in prediction [35,36]. The optimal feature selection varies both with the particular classification problem (e.g., separating two different types of variable stars) and the algorithm used. We have performed an extensive set of experiments for this optimization.

We have employed different classifiers in the selected feature space to assess the performance of different feature selection algorithms, to prove that feature selection strategies actually help in reduce the dimensionality of the problem without loss in accuracy. The performance of the classifiers is rated based on the following three criteria. *Completeness* is the percentage of objects of a given class correctly classified as such. *Contamination* is the percentage of objects of a given class, incorrectly classified as belonging to another class. *Loss* is the fraction of misclassified data.

One potential problem is overfitting, when a data model is excessively complex and starts to fit the noise, instead of the data, thus leading to a deteriorated performance. To avoid it, a cross-validation approach is recommended, e.g., with 10-fold cross-validation the original sample is randomly partitioned into 10 subsamples. Each time a single subsample is retained as test data, and the remaining are used as training data. This process is then repeated 10 times with each of the subsamples used exactly once as the test. In presence of few training data, Leave-One-Out Cross-Validation (LOOCV) may be used: a single observation from the original sample is used as the validation data, and the remaining observations as the training data. This is repeated such that each observation in the sample is used once as the validation data. Leave-one-out cross-validation is usually very computationally expensive because of the large number of times the training process is repeated.

For example in an experiment where we classify two types of variable stars, RR Lyrae and W UMa, using a Relief method, only four parameters out of the 60 available were selected (Fig. 4). For this particular problem, we obtain completeness rates of ~ 96-97%, and contamination rates of ~ 3-4%. In at least some cases the parameters automatically selected by this procedure correspond to physically meaningful relations based on the domain knowledge, even though no such external information was provided to the algorithm. One example is the so-called period-amplitude relationship for a particular type of variable stars (RR Lyrae); the details and the astronomical background to which we refer are beyond the scope of this paper, but see, e.g., [45].

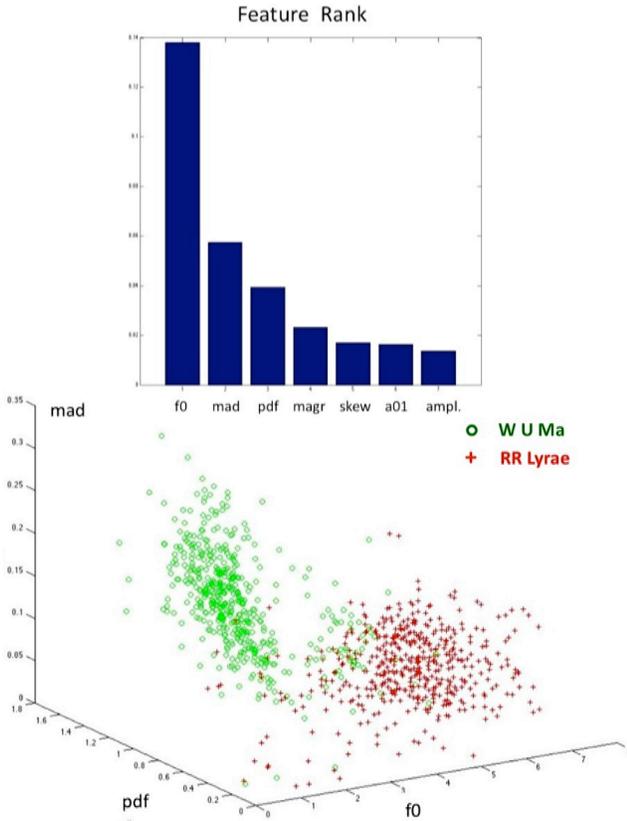

**Figure 4.** Top: A relative ranking of light curve features for a particular classification problem in separating two types of variable stars, RR Lyrae (red crosses) and W UMa (green circles). Bottom: The two classes are separated very effectively in the 3-dimensional space of the 3 top ranked features. The features are defined in [27].

A more challenging, but more realistic and relevant problem is multi-class classification. To find the parameters that give the most of the classification discriminating information, we have used a subset from CRTS containing six classes (Supernovae, Cataclysmic Variables, Blazars, other AGNs, RR Lyrae and Flare Stars) and 20 parameters. Table 1 shows some of these results for two different multi-class experiments. It is interesting to note that different features appear among the most significant subsets, depending on the physical nature of the classes considered.

Thus, we see that feature selection strategies can lead to a substantial dimensionality reduction and improved classifier performance in a broad range of astrophysical situations.

| Class | Compl. | Contam. | Features |
|---|---|---|---|
| Blazar | 81% | 13% | Amplitude, linear_trend, flux_percentile_ratio_mid20, percent_diff_flux_percentile, qso, skew, std, Stetson_j, Stetson_k, lomb-scargle |
| CV | 96% | 5% | |
| RR Lyr | 97% | 5% | |
| SN Ia | 99% | 1% | |

| Class | Compl. | Contam. | Features |
|---|---|---|---|
| Blazar | 83% | 13% | Amplitude, beyond1std, flux_percentile_ratio_mid65, max_slope, qso, std, lomb-scargle |
| CV | 94% | 6% | |
| RR Lyr | 97% | 4% | |

**Table 1:** Completeness and Contamination for two classification experiments separating between four (top) and three different classes (bottom). The features are defined in [27].

IV. NOVEL APPROACHES TO THE PERIOD FINDING

In many situations (not just in astronomy), time series represent periodically variable phenomena. The two primary challenges then are finding out if the signal is periodic, and if so, determining the correct period. This is a serious challenge in the situations where data are severely undersampled and/or irregularly sampled, so that the standard Fourier-based techniques are generally not applicable. Moreover, the waveforms may have an arbitrary shape that is not well fit by a sufficiently small collection of sinusoidal waves.

For example, in astronomy, substantial number of variable stars are periodic in nature, and common problem in time domain astronomy is an optimal determination of their periods from sparse or otherwise limited data. Given the importance of these types of stars for the studies of the Galactic structure (notably the RR Lyrae), the cosmological distance scale (Cepheids and RR Lyrae), and stellar structure and evolution in general, this problem has a strong practical significance.

Synoptic sky surveys now generate literally hundreds of millions of light curves, and many show a statistically significant variability. One of the first questions asked is: could this be a periodic variable star, and if so, what is its period, and the correctly phase-folded light curve? In many cases, that determines the physical nature of the source, and its possible astrophysical uses, since different types of periodic variables often have well defined ranges of periods, amplitudes, or even light curve shapes.

The sheer data volume requires use of automated period-finding algorithms, but different ones work better in different situations, and may produce spurious answers in some cases. For example, the Lomb-Scargle (LS) algorithm [46] is known to commonly misidentify half the period of eclipsing binary stars as the true period, requiring a systematic correction [47]. Periodic signals with sawtooth waveforms or eccentric spectroscopic binary radial velocity curves are also better analyzed with a Hoeffding-test based technique [48]; and so on. In the cases where different algorithms disagree, and/or a statistical measure like the $\chi^2$ indicates a poor fit, in some cases a human disambiguation may be needed, which may require a judicious ignoring of apparent measurement glitches that may be confusing some of the algorithms. As we already described earlier, light curves can be parametrized by a number of statistical descriptors, forming a feature vector, with reliably classified ones providing the labels.

We conducted an extensive comparison study of several of the popular period finding algorithms, and some new ones [29,30] using three different large data sets from three different sky surveys, with different sampling strategies, cadences, S/N, etc. We analyzed the accuracy of the methods against magnitude, sampling rates, quoted period, quality measures (signal-to-noise and number of observations), variability, and object classes. On the whole, we find that measure of dispersion-based techniques (e.g., analysis-of-variance with harmonics, and conditional entropy method) consistently give the best results but there are clear dependencies on object class and light curve quality.

We find that all methods are dependent on the quality of the light curve and show a decline in period recovery with lower quality light curves as a consequence of fewer observations, fainter magnitudes and/or noisier data and an increase in period recovery with higher object variability. This is qualitatively as expected, but we address it in a quantitative detail. Using real data with irregular sampling and heteroskedastic errors, we found generally that period finding algorithms can recover the period of a regularly periodic object with a reasonable degree of accuracy at best 50% of the time. If the focus is only on identifying periodic behavior then rates of ~70% are achievable. One of the issues is that (new) algorithms are often tested on simulated periodic signals, typically sinusoids with Gaussian noise, which do not reflect observed data. Period aliasing and identifying a period harmonic also remain significant issues. We also considered a simple ensemble approach and find that it performs no better than individual algorithms.

Phase dispersion measures are more robust and typically faster than those which fit a basis set of periodic functions, e.g., as in the Fourier decomposition. Pulsating and eclipsing variable classes were also found to be easily identifiable than eruptive, rotating, or other types of periodic object. A bimodal observing strategy consisting of pair of short observations per night allowed better period recovery than just a single visit.

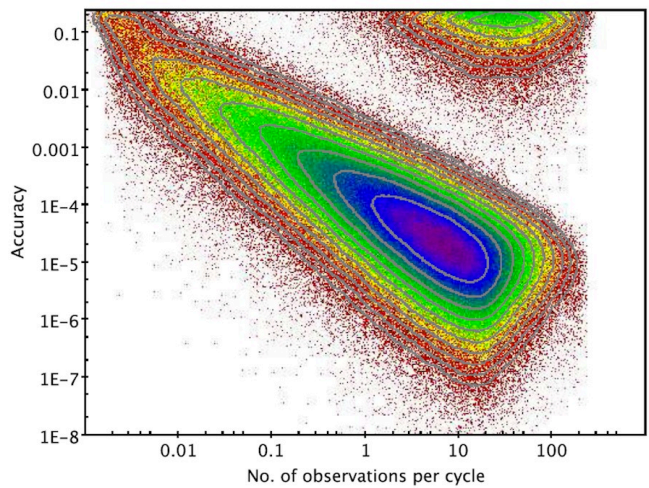

**Figure 5.** An evaluation of the performance of the conditional entropy algorithm [30], using a Monte-Carlo simulation of realistic light curves. The distribution of period determination accuracies from the synthetic data in terms of the number of observations per cycle for a given synthetic light curve. We find that the optimal performance is in a regime where there are ~ 3 to 10 observations per cycle.

We considered the performance of the individual algorithms and show that a new conditional entropy-based algorithm [30], is the most optimal in terms of completeness and speed, as evaluated through extensive Monte-Carlo simulations (Fig. 5). More details are given in that paper.

More recently, we developed some novel algorithms for a detection of a periodic behavior in poorly and irregularly sampled time series, and applied them to the time series of flux measurements of quasars (luminous objects powered by accretion onto supermassive black holes). Quasar variability is known to be stochastic, and until recently there was only one possible case of a periodic behavior.

Wavelets are an increasingly popular tool in analyses of time series and are particularly attractive since they allow both localized time and frequency analysis. In particular, the power spectrum of a time series can be evaluated as a function of time and the time evolution of parameters associated with possible (quasi-)periodic behavior determined, i.e., period, amplitude and phase. Although conventional wavelet analysis via the discrete wavelet transform requires regularly sampled data, a number of techniques have been developed to deal with irregularly sampled data.

We applied the weighted wavelet transform [38] and the z-transform discrete correlation function (ZDCF) [39] to the CRTS light curves of ~ 247,000 known, spectroscopically confirmed quasars [40,41]. Both of these algorithms can detect (quasi-)periodic behavior in irregularly sampled data. We define the period of the quasar from the largest peak in the ZDCF between the second and third zero-crossings of a Gaussian process model fit to the ZDCF [42]. This produced an important astrophysical result, a detection of a periodic behavior in ~ 100 observed quasars [40,41]. The physical interpretation of this involves binary supermassive black holes on their way towards a merger, a process long predicted by theory, but never previously observed.

## V. BEYOND UNSUPERVISED CLUSTERING: MACHINE-ASSISTED DISCOVERY

As the exponential growth of data volumes, rates, and complexity continues, we may see an increased use of methods for a collaborative human-computer discovery. Recognizing meaningful patterns and correlations in high dimensionality data parameter spaces is a very non-trivial task.

Another novel approach that we explored in the course of this study is the use of Machine Discovery, i.e., software that can formulate and test data models. The particular package that we used, with M. Graham as the lead, is $Eureqa^{TM}$ [31]. Here we outline some of the key results; more details are given in [32].

$Eureqa^{TM}$ is a software tool which aims to describe a data set by identifying the simplest mathematical formulae which could describe the underlying mechanism that produced the data. It employs symbolic regression to search the space of mathematical expressions to determine the best-fitting functional form – this involves fitting both the form of the equation and its parameters simultaneously. Binary classification can be cast as a problem amenable to this tool – the "trick" is to formulate the search relationship as: $class = g(f(x_1, x_2, x_3, ..., x_n))$ where $g$ is either the Heaviside step function or the logistic function, which gives a better search gradient. $Eureqa^{TM}$ finds a best-fit function, $f$, to the data that will get mapped to a 0 or a 1, depending on whether it is positively or negatively valued (or lies on either side of a specified threshold, say 0.5, in the case of the logistic function.)

We considered three specific binary light curve classification problems using $Eureqa^{TM}$: RR Lyrae vs. W UMa (Fig. 6), CV vs. blazar, and Type Ia vs. core-collapse Supernovae. For each case, we compiled data sets of light curves from the CRTS survey for the appropriate classes of objects, and derived ~30 – 60 dimensional feature vectors for each object. A set of 10 $Eureqa^{TM}$ runs was performed for each case with each run omitting 10% of the data and the best-fit solution for that run then applied with the omitted data as the validation set so giving us 10x-cross-validation on the resulting solutions.

For example, in the binary classification of these periodic variables, $Eureqa^{TM}$ correctly identifies the optimal feature parameter plane that separates them as physically distinct classes (Fig. 7). This is very impressive, since the program does not "know" anything about these objects, and simply discovers the relationship contained in the data.

Some of the preliminary results for multiple classes, comparing $Eureqa^{TM}$ with one of the best "traditional" machine learning methods, Decision Trees (DT), are given in Table 2. We note that DT is a supervised classification method, and thus it incorporates the domain knowledge from the training data set; $Eureqa^{TM}$ has no such expert-provided input. Even so, the results are broadly comparable for most classes. $Eureqa^{TM}$ does not do as well in the situation where the light curves are qualitatively similar, e.g., blazars vs. CVs, or different subtypes of Supernovae. However, a random person with no expert knowledge in this field (just as *Eureqa*[TM] doesn't have it) would probably also fail completely in separating those classes.

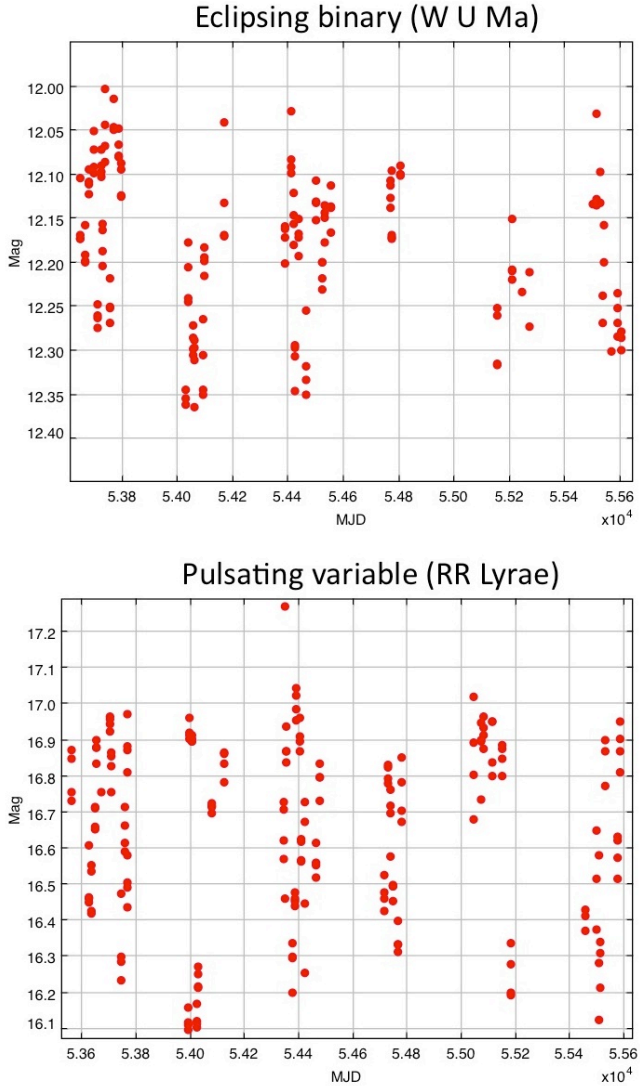

**Figure 6.** An illustration of the classification challenge. Light curves of two types of periodic variables (not folded by the period) from the CRTS survey, used in this experiment. It is not easy to tell that these correspond to two entirely different types of variable stars.

As these preliminary results show, at least in some cases *Eureqa*[TM] can identify and characterize physically meaningful structures in feature vector data to a sufficient degree that it can be employed for binary classification. An advantage of this is that *Eureqa*[TM] provides an analytical expression to separate the classes rather than relying on application of a trained black box algorithm. We see this as one of the first steps in a practical *human-computer collaborative discovery* in the era of big data. We think that such novel methods will become increasingly important for the data-intensive science in the 21$^{st}$ century.

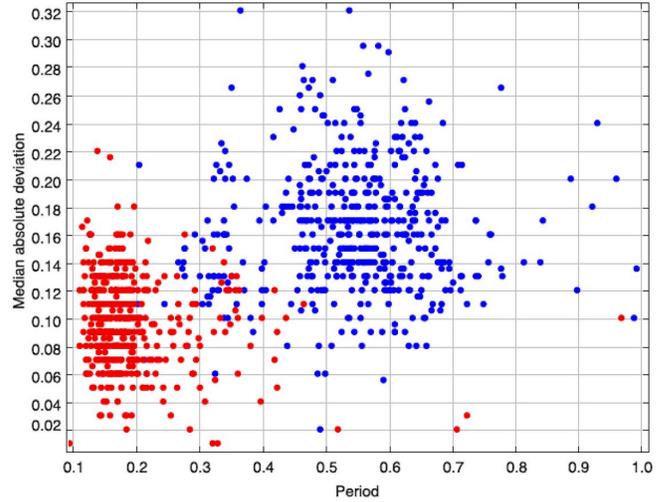

**Figure 7.** Separation of two types of periodic variables, RR Lyrae (blue) and W UMa (red) in the optimal feature plane discovered by *Eureqa*[TM].

|  | *Eureqa*[TM] | | Decision Tree | |
|---|---|---|---|---|
| *Data set* | *Purity* | *Complet.* | *Purity* | *Complet.* |
| RR Lyr | 98% | 96% | 95% | 95% |
| W UMa | 97% | 99% | 96% | 96% |
| CV | 89% | 91% | 92% | 92% |
| Blazar | 68% | 63% | 87% | 83% |
| SN Ia | 76% | 93% | 90% | 96% |
| CC SN | 74% | 41% | 92% | 80% |

**Table 2:** A comparison of the results using *Eureqa*[TM] and a traditional supervised classifier for six different types of variable objects. From [32].

VI. METACLASSIFICATION: OPTIMAL COMBINING OF CLASSIFIERS AND CONTEXTUAL KNOWLEDGE

Direct measurements of various parameters for a given object can be supplemented by the contextual information, e.g., what is around in the image, how did it behave in the past, was it detected on other wavelengths, etc. Such contextual information can be highly relevant to resolving competing interpretations: for example, the light curve and observed properties of a transient might be consistent with both it being a cataclysmic variable star, an active galactic nucleus, or a supernova. If it is subsequently known that there is a galaxy in close proximity, the supernova interpretation becomes much more plausible. Such information, however, can be characterized by high uncertainty and absence, and by a rich structure – if there were two candidate host galaxies, their morphologies, distance, etc., become important, e.g., is this type of supernova more consistent with being in the extended halo of a large spiral galaxy or in close proximity to a faint dwarf galaxy? The ability to incorporate such contextual information in a quantifiable fashion is highly desirable.

We are investigating the use of crowdsourcing as a means of harvesting human pattern recognition skills, especially in the context of capturing the relevant contextual information, and turning it into machine-processable algorithms.

We can identify three possible sources of information that can be used to find the unknown parameters. They can be from *a priori* knowledge, e.g. from physics or monotonicity considerations (e.g., Supernova light curves past the maximum are always declining), or from examples that are labeled by experts, or from the feedback from downstream observatories once labels are determined. The first case would serve to give an analytical form for the distribution, but the second two amount to the provision of labeled examples, $(x, y)$, which can be used to select a set of $k$ probability distributions.

A methodology employing contextual knowledge forms a natural extension to the logistic regression and classification methods mentioned above. Ideally such knowledge can be expressed in a manipulable fashion within a sound logical model, for example, it should be possible to state the rule that "a supernova has a stellar progenitor and will be substantially brighter than it by several order of magnitude" with some metric of certainty and infer the probabilities of observed data matching it.

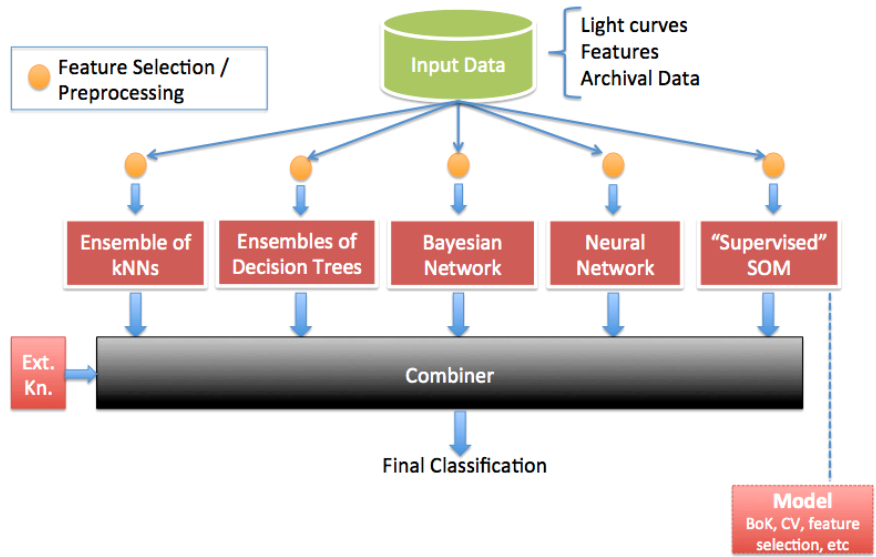

**Figure 8.** A schematic illustration of the metaclassifier for an optimal combination of the output of different classifiers.

In general, we are dealing with a set or classifiers whose outputs may not match, which are also optimized for different parts of the classification problem. For example, one classifier may be very good at separating Supernovae from all other transients, but perform poorly on the classification of different types of variable stars, etc. This leads us to a problem of metaclassification, i.e.,an optimal combining of the outputs of different classifiers with an inclusion of the external domain knowledge and contextual information, illustrated schematically in Fig. 8.

Markov Logic Networks (MLN) [33] are one such probabilistic framework using declarative statements (in the form of logical formulae) as atoms associated with real-valued weights expressing their strength. The higher the weight, the greater the difference in log probability between a world that satisfies the formula and one that does not, all other things being equal. In this way, it becomes possible to specify 'soft' rules that are likely to hold in the domain, but subject to exceptions - contextual relationships that are likely to hold such as supernovae may be associated with a nearby galaxy or objects closer to the Galactic plane may be stars.

The structure of a MLN – the set of formulae with their respective weights – is also not static but can be revised or extended with new formulae either learned from data or provided by third parties. In this way, new information can easily be incorporated. Continuous quantities, which form much of astronomical measurements, can also be easily handled with a hybrid MLN. This approach could be used to represent a set of different classifiers and the inferred most probable state of the world from the MLN would then give the optimal classification.

We are also experimenting with the "sleeping expert" method [34]. A set of different classifiers each generally works best with certain kinds of inputs. Activating these optionally only when those inputs are present provides an optimal solution to the fusion of these classifiers. Sleeping expert can be seen as a generalization of the *if-then* rule: *if* this condition is satisfied *then* activate this expert, e.g., a specialist that makes a prediction only when the instance to be predicted falls within their area of expertise. For example, some classifiers work better when certain inputs are present, and some work only when certain inputs are present. It has been shown that this is a powerful way to decompose a complex classification problem. External or *a priori* knowledge can be used to awake or put experts to sleep and to modify online the weights associated to a given classifier; this contextual information may be also expressed in text.

## VII. CLASSIFICATION-INFORMED AUTOMATED DECISION MAKING

While at least preliminary astrophysical classifications of variable sources and transient events may be obtained using survey and archival data and the methods described above, in many cases the classifications will be ambiguous, or, in the case of particularly interesting events, additional data from other instruments would be needed to fully exploit them scientifically. This poses the challenge of automated decision making as to the optimal use of the available, finite follow-up resources, e.g., other telescopes or instruments. The process is dynamical and iterative, as illustrated schematically in Fig. 9.

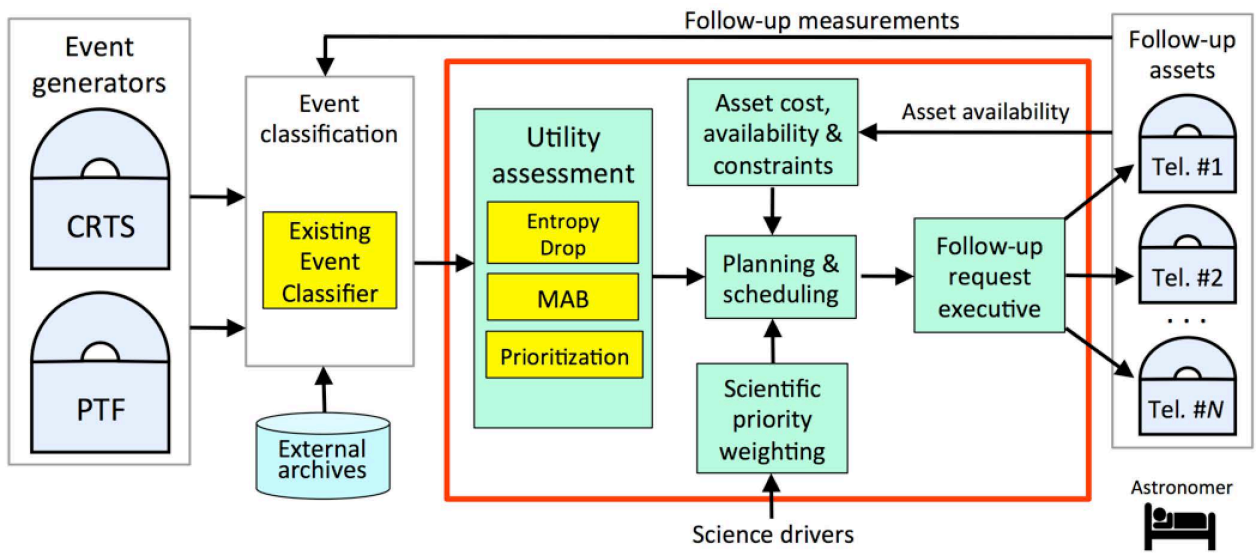

**Figure 9.** A schematic illustration of the overall system. Data from sky survey data streams are fed into the automated event classifiers, supplemented with the archival data resources. This preliminary classification is then fed to the automated decision making engine (outlined as the red box). Utility of various potential follow-up measurements is assessed, and combined with the cost/reward functions based on the availability and cost of follow-up resources, as well as the potential scientific significance of the measured events. Requests for follow-up observations are communicated to a set of participating robotic telescopes, which may obtain additional measurements, and communicate them back to the classifier for an improved classification.

We typically have sparse observations of a given object of interest, leading to classification ambiguities among several possible object types (e.g., when an event is roughly equally likely to belong to two or more possible object classes, or when the initial data are simply inadequate to generate a meaningful classification at all). Generally speaking, some of them would be of a greater scientific interest than others, and thus their follow-up observations would have a higher scientific return. Observational resources are scarce, and always have some cost function associated with them, so a key challenge is to determine the follow-up observations that are most useful for improving classification accuracy, and detect objects of scientific interest.

There are two parts to this challenge. First, what type of a follow-up measurement – given the *available* set of resources (e.g., only some telescopes/instruments may be available) – would yield the maximum information gain in a particular situation? And second, if the resources are finite and have a cost function associated with them (e.g., you can use only so many hours of the telescope time), when is the potential for an interesting discovery worth spending the resources?

This work is still in progress, but we outline here some of the key ideas. We first take an information-theoretic approach to this problem that uses Shannon entropy to measure ambiguity in the current classification. We can compute the entropy drop offered by the available follow-up measurements – for example, the system may decide that obtaining an optical light curve with a particular temporal cadence would discriminate between a supernova and a flaring blazar, or that a particular color measurement would discriminate between, say, a cataclysmic variable eruption and a gravitational microlensing event. A suitable prioritized request for the best follow-up observations would be sent to the appropriate robotic (or even human-operated) telescopes.

Alternatively, instead of maximizing the classification accuracy, we consider a scenario where the algorithm chooses a set of events for follow-up and subsequent display to an astronomer. The astronomer then provides information on how interesting the observation is. The goal of the algorithm is to learn to choose follow-up observations which are considered most interesting. This problem can be naturally modeled using *Multi-Armed Bandit* algorithms (MAB). The MAB problem can abstractly be described as a slot machine with $k$ levers, each of which has different expected returns (unknown to the decision maker). The aim is to determine the best strategy to maximize returns. There are two extreme approaches: (1) exploitation – keep pulling the lever which, as per your current knowledge, returns most, and (2) exploration – experiment with different levers in order to gather information about the expected returns associated with each lever. They key challenge is to trade off exploration and exploitation. There are algorithms guaranteed to determine the best choice as the number of available tries goes to infinity.

In this analogy different telescopes and instruments are the levers that can be pulled. Their ability to discriminate between object classes forms the returns. This works best when the priors are well assembled and a lot is already known about the type of object one is dealing with. But due to the heterogeneity of objects, and increasing depth leading to transients being detected at fainter levels, and more examples of relatively rarer subclasses coming to light, treating the follow-up telescopes as a MAB will provide a useful way to rapidly improve the classification and gather more diverse priors. An analogy could be that of a genetic algorithm which does not get stuck in a local maxima because of its ability to sample a larger part of the parameter space.

VIII. CONCLUDING COMMENTS

Our goal in this paper was to illustrate the richness and the challenges associated with the problem of an automated classification of transient events and variable sources (or, more generally, heterogeneous time series of measurements of a population of objects containing a number of different classes). Whereas this is one of the core challenges of the vibrant and emerging field of time-domain astronomy, similar problems can be easily identified in other domains.

Several aspects of this problem make it particularly interesting: dealing with the data heterogeneity and sparsity; use of statistical descriptors to form feature vectors, instead of using the data directly; dimensionality reduction of feature spaces that is context-dependent; forays into the collaborative human-computer discovery; optimal combining of different classifiers that is also context dependent; and finally, optimal allocation of limited follow-up resources when there are multiple cost functions involved.

ACKNOWLEDGMENT

This work was supported in part by the NASA grant 08-AISR08-0085, the NSF grants AST-0909182, IIS-1118041, and AST-1313422, by the W. M. Keck Institute for Space Studies at Caltech (KISS), and by the U.S. Virtual Astronomical Observatory, itself supported by the NSF grant AST-0834235. Some of this work was assisted by the Caltech students Nihar Sharma, Yutong Chen, Alex Ball, Victor Duan, Allison Maker, and others, supported by the Caltech SURF program. We thank numerous collaborators and colleagues, especially within the CRTS survey team, and the world-wide Virtual Observatory and astroinformatics community, for stimulating discussions.